\documentclass[showpacs,twocolumn,preprintnumbers,showkeys,superscriptaddress,amsmath,amssymb,nofootinbib]{revtex4-1}

\usepackage{lipsum}

\usepackage{bbold}
\usepackage{color}
\usepackage{latexsym}
\usepackage{amsmath}
\usepackage{amssymb}
\usepackage[utf8]{inputenc}
\usepackage{amsfonts}
\usepackage{bm}
\usepackage{bbold}

\usepackage{eufrak}
\usepackage{euscript}
\usepackage{graphics}
\usepackage{graphicx}

\newcommand{\be}{\begin{equation}}
\newcommand{\ee}{\end{equation}}
\newcommand{\ba}{\begin{eqnarray}}
\newcommand{\ea}{\end{eqnarray}}

\begin{document}

%

\title{\Large Gambini-Pullin Electrodynamics as a scenario for Cherenkov radiation in QED vacuum}

\author{Patricio Gaete} \email{patricio.gaete@usm.cl}
\affiliation{Departamento de F\'{i}sica and Centro Cient\'{i}fico-Tecnol\'ogico de Valpara\'{i}so-CCTVal,
Universidad T\'{e}cnica Federico Santa Mar\'{i}a, Valpara\'{i}so, Chile}

\author{J. A. Helay\"el-Neto}\email{helayel@cbpf.br}
\affiliation{Centro Brasileiro de Pesquisas F\'isicas, Rua Dr. Xavier Sigaud
150, Urca, Rio de Janeiro, Brasil, CEP 22290-180}

\date{\today}

\begin{abstract}

We examine the electromagnetic radiation produced by a moving charge in the QED vacuum that behaves as a dispersive medium characterized by a geometrical structure (discreteness/granularity) that emerges from loop quantum gravity. It is shown that the radiation is driven by the refractive vacuum the charged particle travels through reproducing the profile of the Cherenkov effect.
\end{abstract}

\maketitle

\pagestyle{myheadings}
\markright{Gambini-Pullin Electrodynamics as a scenario for Cherenkov radiation in QED vacuum}

\section{Introduction}

Lorentz invariance is a key ingredient of quantum field theory, which is a exact symmetry of the Standard Model (SM) \cite{Colladay} of the fundamental constituents of the matter. As is well known, its great triumph lies in being able to predict with an impressive accuracy every process in the accelerators to this day.  Notwithstanding the above, this success is not complete because the SM does not include gravity, for which we still do not have a quantum description in four space-time dimensions. In this connection, it is assume that these theories will merge at the Planck scale 
${m}_{P}\approxeq{10}^{19}{GeV}$, in order to provide a consistent framework to unify all fundamental interactions \cite{Kostelecky}.

Nevertheless, at these energies Lorentz-and CPT-symmetry violation effects are expected to become manifest, which has generated a great research interest. The motivation for the investigation of such effects is that in some theories that aspire to give a quantum description of gravity, such as string theory \cite{KosteSamuel1, KosteSamuel2,KosteSamuel3,KosteSamuel4} and loop quantum gravity \cite{Gambini,Alfaro,Urrutia}, the breakdown of Lorentz and CPT symmetries symmetries take place. Interestingly, the physical picture which emerges from these scenarios is that there would be atoms of geometry of which fundamental building blocks of gravity arise (discreteness). In other words, the above developments encode the intrinsic property of granularity of a quantum spacetime. We further note that the focus of these studies lies on low-energies, where the effects of Lorentz-symmetry violation (LSV) can be displayed. Indeed, one of the most interesting consequences of the previous discreteness is that it would act as a dispersive medium for particles, such as photons, propagating on this spacetime.

With these considerations in mind, in a previous work \cite{Gaete2019}, we have introduced a new Maxwell electrodynamics coupled to a Lorentz-violating background through the presence of higher-derivative terms. This study was carried out in the familiar language of standard quantum field theory. In fact, the implications of this effective theory modified by Lorentz-violating operators of mass dimension-$5$ were analised,
such as modified dispersion relations which exhibit the vacuum birefringence phenomenon. Once this was done, by computing the energy-momentum tensor we have reproduced the Gambini-Pullin electrodynamics \cite{Gambini}. In particular, it was shown that for a background time-like four-vector, ${n}^{\mathit{\mu}}$, the equations of motion reproduce exactly those obtained by Gambini and Pullin in the framework of loop quantum gravity.

However, in order to put our discussion into context it is useful to recall that the dispersion relation to order ${\cal O}(l_P)$, for Gambini-Pullin electrodynamics, is given by \cite{Gaete2019}:
\begin{equation}
{w}^{2}{=}\,{k}^{2}\,{+}\,{2}\,{l}_{P}\,\mathit{\eta}\,{k}^{3}, \label{I-05}
\end{equation}
where ${k}\,{=}\!\!\mid{\!\!{\bf k}}{\mid}$, $l_P$ is Planck's length and $\mathit{\eta}\,{=}\pm{1}$ is a parameter to describe left-handed and right-handed modes.

Hence we see that the refractive index is given by 
 \begin{equation}
 {n}(w) \!= \!1\! - {l_P}\,\eta\, w, \label{I-10}
 \end{equation}
 in the same way we have that the permittivity ($\varepsilon$) and permeability ($\mu$) are given by $\varepsilon\equiv\mathit{\mu}\,{=}\,{1}{-}\,{l}_{P}\,\mathit{\eta}\,{w}$.

Inspired by these observations, the purpose of this work is to extend our previous study \cite{Cherenkov23} on electromagnetic (Cherenkov) radiation in a medium characterized by nonlinearities of the electromagnetic field. In fact, in the preceding work the permittivity and susceptibility induced by nonlinearities were studied in order to obtain the refractive index for logarithmic electrodynamics. In addition, we computed the electromagnetic radiation produced by a moving charged particle interacting with a medium characterized by nonlinearities of the electromagnetic field. The result showed that the radiation is like the one that happens in the Cherenkov effect.

In this context it is useful to recall that quantum vacuum nonlinearities have a long history originating from the pioneering work by Euler and Heisenberg \cite{EH}, who obtained an effective nonlinear electromagnetic theory in vacuum arising from the interaction of photons with virtual electron-positron pairs. In fact, one of the most startling predictions of the Heisenberg and Euler result has been vacuum birefringence. More precisely, the quantum vacuum when is stressed by external electromagnetic fields behaves as if it were a birefringent material medium, which has been emphasized from different viewpoints \cite{Adler,Costantini,Ruffini,Dunne}.

Since Gambini-Pullin electrodynamics is characterized by a dispersive medium arising of a geometrical structure, therefore it should be natural to ask wether electromagnetic radiation is present in this case.
In the present work, following the standard approach of calculating the Poynting vector,
we attempt to examine this issue. As we will see, our analysis reveals the vital role played by the new geometrical vacuum in triggering the radiated energy.


\section{Electromagnetic radiation}

We turn now to the problem of obtaining the electromagnetic radiation produced by a moving charged particle interacting with a medium characterized by a structure that emerges from loop quantum gravity. In order to put our discussion into context, it is appropriate to start first by considering the new Maxwell equations for a moving charged particle in a medium characterized by Gambini-Pullin electrodynamics:
\begin{eqnarray}
\nabla  \cdot {\bf E} = \frac{{4\pi }}{\varepsilon }{\rho _{ext}}, \nonumber\\
\nabla  \cdot {\bf B} = 0, \nonumber\\
\nabla  \times {\bf E} + {l_P}\nabla  \times \left( {\nabla  \times {\bf E}} \right) =  - \frac{1}{c}\frac{{\partial {\bf B}}}{{\partial t}}, \nonumber\\
\nabla  \times {\bf B} + {l_P}\nabla  \times \left( {\nabla  \times {\bf B}} \right) = \frac{{\varepsilon \mu }}{c}\frac{{\partial {\bf E}}}{{\partial t}} + \frac{{4\pi \mu }}{c}{{\bf J}_{ext}}. \label{II-05}
\end{eqnarray}
Here, ${\rho _{ext}}$ and ${{\bf J}_{ext}}$ denote the external charge and current densities. 

It is of interest also to notice that the foregoing equations at leading order in $l_{P}$ can be written alternatively in the form
\begin{widetext}
\begin{subequations}
\begin{eqnarray}
({\nabla ^2} - \frac{1}{{{c^{\prime 2}}}}\frac{{{\partial ^2}}}{{\partial {t^2}}}){\bf E} + 2{l_P}{\nabla ^2}\left( {\nabla  \times {\bf E}} \right) = 
\frac{{4\pi }}{\varepsilon }\nabla \rho_{ext}  
+ \frac{{4\pi \mu }}{{{c^2}}}\frac{{\partial {{\bf J}_{ext}}}}{{\partial t}}, 
\label{II-10} \\
\left( {{\nabla ^2} -\frac{1}{{{c^{\prime 2}}}}\frac{{{\partial ^2}}}{{\partial {t^2}}}} \right){\bf B} + 2{l_P}{\nabla ^2}\left( {\nabla  \times {\bf B}} \right) = 
 - \frac{{4\pi \mu }}{c}\nabla  \times {\bf J}_{ext} + {l_P}\frac{{4\pi \mu }}{c}\left( {{\nabla ^2}{\bf J}_{ext} + \frac{\partial }{{\partial t}}\nabla \rho_{ext} } \right),\label{II-15}
 \end{eqnarray}
 \end{subequations}
 \end{widetext}
where the external charge and current densities are given by: ${\rho _{ext}}\left( {t,{\bf x}} \right) = Q\delta \left( x \right)\delta \left( y \right)\delta \left( {z - vt} \right)$ and ${\bf J}_{ext}\left( {t,{\bf x}} \right) = Qv\delta \left( x \right)\delta \left( y \right)\delta \left( {z - vt} \right){\hat {\bf e}_z}$.
We mention in passing that, for simplicity, we are considering the $z$ axis as the direction of the moving charged particle.  

In order to solve equations (\ref{II-10}) and (\ref{II-15}), we shall begin by noting that such equations can be written in matrix form as
\begin{equation}
{M}_{ij}\,{v}_{j}\,{=}\,{w}_{i}, \label{II-15a}
\end{equation}
where $v$ stands for the electric and magnetic fields, while $w$ designates the right hand side of the equations for the electric and magnetic fields.
In such a case the matrix, $M_{ij}$, is given by $
{M}_{ij}\equiv{\mathit{\alpha}\,\mathit{\delta}}_{ij}\,{+}\,{\mathit{\beta}\,\varepsilon}_{ilj}{\partial}_{l}$, where $
\mathit{\alpha}\,{=}\,{\nabla}^{2}{-}\frac{1}{{c}^{\prime{2}}}\frac{{\partial}^{2}}{\partial{t}^{2}}$ and $\mathit{\beta}\,{=}\,{2}\,{l}_{P}\,{\nabla}^{2}$. 

From the above it now follows that the inverse matrix takes the form
\begin{equation}
{M}_{jm}^{{-}{1}}\,{=}\,\frac{1}{\det M}\left({{\mathit{\alpha}}^{2}\,{\mathit{\delta}}_{jm}{+}{\mathit{\alpha}\,\mathit{\beta}\varepsilon}_{jmn}{\partial}_{n}{+}{\mathit{\beta}}^{2}\,{\partial}_{j}{\partial}_{m}}\right),  \label{II-15b}   
\end{equation}
where $
\det\hspace{0.33em}\!\!{M}\,{=}\,\mathit{\alpha}\left({{\mathit{\alpha}}^{2}\,{+}\,{\mathit{\beta}}^{2}\,{\nabla}^{2}}\right)$.

Making use of this last equation, we find that the magnetic and electric fields, to order ${\cal O}(l_P)$, become 
\begin{eqnarray}
{\bf B}\left({t,{\bf x}}\right)\!\!&=&\!\!{-}\frac{{4}\mathit{\pi}\mathit{\mu}}{c}\frac{1}{\mathit{\alpha}}  {\nabla} \times{\bf J}_{{ext}\hspace{0.33em}}{+}\,{l}_{P}\frac{{4}\mathit{\pi}\mathit{\mu}}{c}\frac{1}{\mathit{\alpha}}{\nabla}^{2}{\bf J}_{ext} \nonumber\\
&{+}&\!\!{l}_{P}\frac{{4}\mathit{\pi}\mathit{\mu}}{c}\frac{1}{\mathit{\alpha}}\frac{\partial}{\partial{t}}\nabla {\mathit{\rho}}_{ext}\,{-}\,\frac{{4}\mathit{\pi}\mathit{\mu}}{c}\frac{\mathit{\beta}}{{\mathit{\alpha}}^{2}}\nabla \times\left({\nabla \times{\bf J}_{ext}}\right), \nonumber\\
\label{II-15c} 
\end{eqnarray}
and
\begin{eqnarray}
{\bf E}\left({t,{\bf x}}\right)\!\!&{=}\!\!&\frac{{4}\mathit{\pi}}{\varepsilon}\frac{1}{\mathit{\alpha}}\nabla {\mathit{\rho}}_{ext}\,{+}\,\frac{{4}\mathit{\pi}\mathit{\mu}}{{c}^{2}}\frac{1}{\mathit{\alpha}}\frac{\partial}{\partial{t}}{\bf J}_{ext} \nonumber\\
&{+}&\frac{{4}\mathit{\pi}}{\varepsilon}\frac{\mathit{\beta}}{{\mathit{\alpha}}^{2}}\nabla \times\nabla {\mathit{\rho}}_{ext} 
\,{+}\,\frac{{4}\mathit{\pi}\mathit{\mu}}{{c}^{2}}\frac{\mathit{\beta}}{{\mathit{\alpha}}^{2}}\nabla \times\frac{\partial}{\partial{t}}{\bf J}_{ext}. \nonumber\\
\label{II-15d} 
\end{eqnarray}

Next, as we have indicated in \cite{Cherenkov23}, we Fourier transform to momentum space via
\begin{equation}
F(t,{\bf x}) = \int {\frac{{dw{d^3}{\bf k}}}{{{{\left( {2\pi } \right)}^4}}}} {e^{ - iwt + {\bf k} \cdot {\bf x}}}F\left( {w,{\bf k}} \right), \label{II-20}
\end{equation}
where $F$ stands for the electric and magnetic fields. 

Once this is done, we arrive at the following expressions for the magnetic and electric fields: 
\begin{eqnarray}
{\bf B}\left( {w,{\bf k}} \right) &=&  - i\frac{{4\pi \mu }}{c}\frac{{{\bf k} \times {{\bf J}_{ext}}}}{\cal O} + {l_P}\frac{{4\pi \mu }}{c}\frac{{\left( {{{\bf k}^2}{{\bf J}_{ext}} + w{\bf k}{\rho _{ext}}} \right)}}{\cal O} \nonumber\\
&-& {l_P}\frac{{8\pi \mu }}{c}\frac{{{{\bf k}^2}}}{{{{\cal O}^2}}}{\bf k} \times \left( {{\bf k} \times {{\bf J}_{ext}}} \right), \label{II-25}
\end{eqnarray}
and
\begin{eqnarray}
{\bf E}\left( {w,{\bf k}} \right) &=& i\frac{{4\pi }}{\varepsilon }\frac{1}{\cal O}{\bf k}{\rho _{ext}} - \frac{{4\pi \mu }}{{{c^2}}}\frac{1}{\cal O}w{{\bf J}_{ext}} \nonumber\\
&-& {l_P}\frac{{8\pi \mu }}{{{c^2}}}\frac{{{{\bf k}^2}}}{{{{\cal O}^2}}}w\left( {{\bf k} \times {{\bf J}_{ext}}} \right). \label{II-30}
\end{eqnarray}
It should, however, be noted that in the last equations we have changed the notation of $\alpha$ to $\cal O$, which is given by  
\begin{equation}
{\cal O} = \frac{{{w^2}}}{{{c^{\prime 2}}}} - {{\bf k}^2},  \  \  \
\frac{1}{{{c^{\prime 2}}}} \equiv \frac{{\varepsilon \mu }}{{{c^2}}}. \label{II-35}
\end{equation}

In a similar manner the external charge and current densities, in the Fourier space take the form: $\rho_{ext} \left( {w,{\bf k}} \right) = 2\pi Q\delta \left( {w - {k_z}v} \right)$ and ${{\bf J}_{ext}}\left( {w,{\bf k}} \right) = 2\pi Qv\delta \left( {w - {k_z}v} \right){\hat {\bf e}_z}$.

By proceeding in the same way as in \cite{Cherenkov23}, we shall now compute ${\bf B}\left( {w,{\bf x}} \right)$ and ${\bf E}\left( {w,{\bf x}} \right)$. In such a case, ${\bf B}\left( {w,{\bf x}} \right)$ is given by
\begin{equation} 
{\bf B}\left( {w,{\bf x}} \right) = \int {\frac{{{d^3}{\bf k}}}{{{{\left( {2\pi } \right)}^3}}}} \; {e^{i{\bf k} \cdot {\bf x}}}\;{\bf B}\left( {w,{\bf k}} \right). \label{II-40}
\end{equation}

Making use of the axial symmetry of the problem under consideration and using cylindrical coordinates, the magnetic field (\ref{II-40}) reads
\begin{widetext}
\begin{eqnarray}
{\bf B}\left( {w,{\bf x}} \right) =  - i\,\frac{{\mu Q}}{{\pi c}}\,{e^{iw{z \mathord{\left/
 {\vphantom {z v}} \right.
 \kern-\nulldelimiterspace} v}}}\int_0^\infty  {d{k_T}{k_T}} \int_0^{2\pi } {d\alpha } \frac{{{e^{i{k_T}{x_T}\cos \alpha }}}}{{{\cal O}{|_{{k_z} = {w \mathord{\left/
 {\vphantom {w v}} \right.
 \kern-\nulldelimiterspace} v}}}}}\left[ {\left( {{k_T}\,sen\alpha \,\pmb {\hat \rho} - {k_T}\cos \alpha \,\pmb {\hat \phi} } \right) - i{l_P}\,{{\bf k}^2} \, \pmb {{\hat e}_z}} \right] \nonumber\\
+\, {l_P}\frac{{\mu Q}}{{\pi c}}\frac{w}{v}\,{e^{iw{z \mathord{\left/
 {\vphantom {z v}} \right.
 \kern-\nulldelimiterspace} v}}}\int_0^\infty  {d{k_T}{k_T}} \int_0^{2\pi } {d\alpha } \frac{{{e^{i{k_T}{x_T}\cos \alpha }}}}{{{\cal O}{|_{{k_z} = {w \mathord{\left/
 {\vphantom {w v}} \right.
 \kern-\nulldelimiterspace} v}}}}}\left[ {\left( {{k_T}\cos \alpha \, \pmb {\hat \rho}  + {k_T} \,sen\alpha \, \pmb {\hat \phi} } \right) + \frac{w}{v} \,\pmb {{\hat e}_z}} \right] \nonumber\\
- 2\,{l_P}\frac{{\mu Q}}{{\pi c}}\,{e^{iw{z \mathord{\left/
 {\vphantom {z v}} \right.
 \kern-\nulldelimiterspace} v}}}\int_0^\infty  {d{k_T}{k_T}} \int_0^{2\pi } {d\alpha } \frac{{{e^{i{k_T}{x_T}\cos \alpha }}}}{{{{\cal O}^2}{|_{{k_z} = {w \mathord{\left/
 {\vphantom {w v}} \right.
 \kern-\nulldelimiterspace} v}}}}}\left[ {\frac{w}{v}\left( {{k_T}\,\cos \alpha \,\pmb {\hat \rho}  + {k_T}\,sen\alpha \,\pmb {\hat \phi} } \right) - k_T^2 \, \pmb {{\hat e}_z}} \right]{{\bf k}^2}, \label{II-45}
\end{eqnarray}
\end{widetext}
where $\pmb{\hat \rho}$ and $\pmb{\hat \phi}$ are unit vectors normal and tangencial to the cylindrical surface, respectively. While $\pmb {{{\hat e}_z}}$ is a unit vector along the $z$ direction. Here, the subscript $T$ in $k_{T}$ simply indicates transversal to the $z$ direction. We also notice that ${\left. {\cal O} \right|_{{k_z} = {w \mathord{\left/
 {\vphantom {w v}} \right.
 \kern-\nulldelimiterspace} v}}} = {w^2}\left( {\frac{1}{{{c^{\prime 2}}}} - \frac{1}{{{v^2}}}} \right) - {\bf k}_T^2$.

It may be observed here that $\int_0^{2\pi } {d\theta } {e^{ix\cos \theta }}\sin \theta  = 0$, $\int_0^{2\pi } {d\theta } {e^{ix\cos \theta }}\cos \theta  = 2\pi i{J_1}\left( x \right)$ and $\int_0^{2\pi } {d\theta } {e^{ix\cos \theta }} = 2\pi {J_0}\left( x \right)$, where ${{J_0}\left( x \right)}$ and ${{J_1}\left( x \right)}$ are Bessel functions of the first kind. With the aid of the previous integrals, we find that  equation (\ref{II-45}) becomes
\begin{widetext}
\begin{eqnarray}
 {\bf B}\left( {w,{\bf x}} \right) = \frac{{2\mu Q}}{c}{e^{iw{z \mathord{\left/
 {\vphantom {z v}} \right.
 \kern-\nulldelimiterspace} v}}} 
 \int_0^\infty  {d{k_T}{k_T}} \left[ { - {k_T}\frac{{{J_1}\left( {{k_T}{x_T}} \right)}}{{{\cal O}{|_{{k_z} = {w \mathord{\left/
 {\vphantom {w v}} \right.
 \kern-\nulldelimiterspace} v}}}}} \pmb {\hat \phi}  - {l_P}{{\bf k}^2}\frac{{{J_0}\left( {{k_T}{x_T}} \right)}}{{{\cal O}{|_{{k_z} = {w \mathord{\left/
 {\vphantom {w v}} \right.
 \kern-\nulldelimiterspace} v}}}}} \pmb {{\hat e}_z}} \right] \nonumber\\
 + \frac{{2{l_P}\mu Q}}{c}{e^{iw{z \mathord{\left/
 {\vphantom {z v}} \right.
 \kern-\nulldelimiterspace} v}}}\int_0^\infty  {d{k_T}{k_T}} \left[ {i\frac{w}{v}{k_T}\frac{{{J_1}\left( {{k_T}{x_T}} \right)}}{{{\cal O}{|_{{k_z} = {w \mathord{\left/
 {\vphantom {w v}} \right.
 \kern-\nulldelimiterspace} v}}}}} \pmb {\hat \rho}  + \frac{{{w^2}}}{{{v^2}}}\frac{{{J_0}\left( {{k_T}{x_T}} \right)}}{{{\cal O}{|_{{k_z} = {w \mathord{\left/
 {\vphantom {w v}} \right.
 \kern-\nulldelimiterspace} v}}}}} \pmb {{\hat e}_z}} \right] \nonumber\\
- \frac{{4{l_P}\mu Q}}{c}{e^{iw{z \mathord{\left/
 {\vphantom {z v}} \right.
 \kern-\nulldelimiterspace} v}}}\int_0^\infty  {d{k_T}{k_T}} \left[ {i\frac{w}{v}{k_T}\frac{{{J_1}\left( {{k_T}{x_T}} \right)}}{{{{\cal O}^2}{|_{{k_z} = {w \mathord{\left/
 {\vphantom {w v}} \right.
 \kern-\nulldelimiterspace} v}}}}} \pmb {\hat \rho}  - k_T^2\frac{{{J_0}\left( {{k_T}{x_T}} \right)}}{{{{\cal O}^2}{|_{{k_z} = {w \mathord{\left/
 {\vphantom {w v}} \right.
 \kern-\nulldelimiterspace} v}}}}} \pmb {{\hat e}_z}} \right]{{\bf k}^2}. \label{II-50}
\end{eqnarray}
\end{widetext} 
 
Then, by integrating over $k_T$ and performing further manipulations, we find that equation (\ref{II-50}) may be rewritten as
 \begin{widetext}
\begin{eqnarray}
{\bf B}\left( {w,{\bf x}} \right)\!\! &=& \!\! - 2i{l_P}\frac{{\mu Qw}}{{cv}}{e^{iw{z \mathord{\left/
 {\vphantom {z v}} \right.
 \kern-\nulldelimiterspace} v}}} 
\left[ {3\sigma {K_1}\left( {\sigma {x_T}} \right) \!-\! {x_T}\left( {{\sigma ^2} - \frac{{{w^2}}}{{{v^2}}}} \right)\!{K_0}\left( {\sigma {x_T}} \right)} \right] \pmb  {\hat \rho} 
+ 2\frac{{\mu Q}}{c}{e^{iw{z \mathord{\left/
 {\vphantom {z v}} \right.
 \kern-\nulldelimiterspace} v}}}\sigma {K_1}\left( {\sigma {x_T}} \right) \pmb {\hat \phi} \nonumber\\
&-&\!\!
{4}{l}_{P}\frac{\mathit{\mu}{Q}}{c}{e}^{{iwz}\slash{v}}\left[{\left({{-}{3}{\mathit{\sigma}}^{2}{+}\frac{{w}^{2}}{{v}^{2}}}\right){K}_{0}\left({\mathit{\sigma}{x}_{T}}\right){+}\frac{\mathit{\sigma}{x}_{T}}{2}\left({{\mathit{\sigma}}^{2}{-}\frac{{w}^{2}}{{v}^{2}}}\right){K}_{1}\left({\mathit{\sigma}{x}_{T}}\right)}\right] \pmb {{\hat e}_z}, 
\label{II-55}
\end{eqnarray}
\end{widetext} 
where ${\sigma ^2} = {w^2}\left( {\frac{1}{{{v^2}}} - \frac{1}{{{c^{ \prime 2}}}}} \right)$. In the above we have used ${x_T} = \rho$ (in cylindrical coordinates). We also have that
${{K_0}\left( x \right)}$ and ${{K_1}\left( x \right)}$ are modified Bessel functions.
 
Our next objective is to compute the electric field. Once again following the same procedure we performed in the above case, we find, from the expression (\ref{II-30}), that the electric field reduces to 
\begin{widetext}
\begin{eqnarray}
{\bf E}\left( {w,{\bf x}} \right) &=& i\frac{{2Q}}{{\varepsilon v}}{e^{iw{z \mathord{\left/
 {\vphantom {z v}} \right.
 \kern-\nulldelimiterspace} v}}} 
\int_0^\infty \!\! {d{k_T}\,{k_T}\!\left( {i{k_T}\frac{{{J_1}\left( {{k_T}{x_T}} \right)}}{{{\cal O}{|_{{k_z} = {z \mathord{\left/
 {\vphantom {z v}} \right.
 \kern-\nulldelimiterspace} v}}}}} \pmb  {\hat \rho}  + \frac{w}{v}\frac{{{J_0}\left( {{k_T}{x_T}} \right)}}{{{\cal O}{|_{{k_z} = {w \mathord{\left/
 {\vphantom {w v}} \right.
 \kern-\nulldelimiterspace} v}}}}} \pmb {{\hat e}_z}} \right)}  
+ i\frac{{2\mu Q}}{{{c^2}}}{e^{iw{z \mathord{\left/
 {\vphantom {z v}} \right.
 \kern-\nulldelimiterspace} v}}} 
\int_0^\infty  {d{k_T}{k_T}w\frac{{{J_0}\left( {{k_T}{x_T}} \right)}}{{{\cal O}{|_{{k_z} = {w \mathord{\left/
 {\vphantom {w v}} \right.
 \kern-\nulldelimiterspace} v}}}}}} \pmb {\hat e_z} \nonumber\\
&+& i{l_P}\frac{{4\mu Q}}{{{c^2}}}{e^{iw{z \mathord{\left/
 {\vphantom {z v}} \right.
 \kern-\nulldelimiterspace} v}}}\int_0^\infty  {d{k_T}k_T^2w\left( {k_T^2 + \frac{{{w^2}}}{{{v^2}}}} \right)\frac{{{J_1}\left( {{k_T}{x_T}} \right)}}{{{{\cal O}^2}{|_{{k_z} = {w \mathord{\left/
 {\vphantom {w v}} \right.
 \kern-\nulldelimiterspace} v}}}}}} \pmb{\hat \phi}. 
\label{II-60}
\end{eqnarray}
\end{widetext}

Again, carrying out the integral over $k_T$, equation (\ref{II-60}) can therefore be written as follows 
\begin{widetext}
\begin{eqnarray}
{\bf E}\left( {w,{\bf x}} \right)\!\! &=&\!\! \frac{{2Q}}{{\varepsilon v}}{e^{iw{z \mathord{\left/
 {\vphantom {z v}} \right.
 \kern-\nulldelimiterspace} v}}}\sigma {K_1}\left( {\sigma {x_T}} \right) \pmb {\hat \rho} \nonumber\\
&+&\!\! i{l_P}\frac{{4\mu Q}}{{{c^2}}}{e^{iw{z \mathord{\left/
 {\vphantom {z v}} \right.
 \kern-\nulldelimiterspace} v}}}w 
\left[ {\sigma {K_1}\left( {\sigma {x_T}} \right) - \frac{{{x_T}}}{2}\left( {{\sigma ^2} - \frac{{{w^2}}}{{{v^2}}}} \right){K_0}\left( {\sigma {x_T}} \right)} \right] \pmb {\hat \phi} \nonumber\\
&-&\!\! i\frac{{2Q}}{{\varepsilon {v^2}}}{e^{iw{z \mathord{\left/
 {\vphantom {z v}} \right.
 \kern-\nulldelimiterspace} v}}}w\left( {1 - \frac{{{v^2}}}{{{c^{ * 2}}}}} \right){K_0}\left( {\sigma {x_T}} \right) \pmb {\hat e_z}.
\label{II-65}
\end{eqnarray}
\end{widetext}

With the foregoing information, we can now proceed to obtain the radiated energy in the case under consideration, our analysis follows closely that of \cite{Cherenkov23}. As already mentioned, in order to accomplish this purpose, we will estimate the radiated energy by calculating the Poynting vector.

It is appropriate to start first by recalling that the density of power carried out by the radiation fields across the surface $S$ bounding the volume $V$ is given by the real part of the Poynting vector (time averaged value)
\begin{equation}
{\bf S} = \frac{c}{{2\pi }}{\mathop{\rm Re}\nolimits} \left[ {\left( {{\bf E} \times {{\bf B}^ * }} \right) + {l_P}\nabla \left( {{\bf E} \cdot {{\bf B}^ * }} \right)} \right]. \label{II-70}
\end{equation}
Note the presence of the last term on the right-hand side which depends on $l_P$.

To be more precise, we will compute the power radiated through the surface $S$ \cite{Das}, which reads
 \begin{equation}
{\cal E} = \int_{ - \infty }^\infty  {dt} \int\limits_S {d{\bf a} \cdot {\bf S}}.  \label{II-75}
\end{equation}
In passing we also note that in our calculation we shall consider a cylinder of radius ${\rho _0}$ (infinitesimally small) as the integration surface \cite{PRA}. 

With this in view, in cylindrical coordinates, the power radiated per unit length through the surface $S$ may be written in the form 
\begin{eqnarray}
{\cal E} &=& \frac{c}{{2\pi }}{\mathop{\rm Re}\nolimits} \int_0^\infty  {dw} \left\{ {2\pi {\rho _0}{{\left. {{S_\rho }} \right|}_{\rho  = {\rho _0}}}} \right\}  \nonumber\\
&+& \frac{c}{{2\pi }}{\mathop{\rm Re}\nolimits} \int_0^\infty  {dw} \left\{ {\frac{\partial }{{{\partial _z}}}\int_0^{{\rho _0}} {\int_0^{2\pi } {{S_z}} \,\rho\, d\rho\, d\phi } } \right\}, \label{II-80}
\end{eqnarray}
in the limit ${\rho _0} \to 0$. It should be noted that in this expression  the $\phi$-component of the Poynting vector $\left( {{S_\phi }} \right)$ does not contribute to the radiated energy.

Now, by employing equations (\ref{II-55}) and (\ref{II-65}), the expression for the power radiated per unit length (\ref{II-80}) reduces to
\begin{equation}
{\cal E} = \frac{{2\pi {Q^2}}}{{{c^2}}}\int_0^\infty  {dw} \, {n^2}(w)\,w\left( {1 - \frac{{{c^2}}}{{{n^2}(w)\,{v^2}}}} \right), \label{II-85}
\end{equation}
where ${n^2}(w) \!= \!1\! - 2\,{l_P}\,\eta\, w$. In this last line we have used the asymptotic behavior of the Bessel function (${K_\nu }\left( x \right) \to \frac{\pi }{{\sqrt {2x} }}{e^{ - x}}$), since we are describing outgoing radiation. It should, however, be noted here that this last expression is similar to that obtained in the Cherenkov radiation theory \cite{Das}, except for an extra ${n^2}(w)$ in the integrand of equation (\ref{II-85}). In fact, this extra ${n^2}(w)$ contributes for corrections of order $l_P$ to the radiated energy.

Before we proceed further, we should comment on our result.
 In the first place and as is well known, ${\cal E}$, represents the rate of energy lost due to radiation along the trajectory of the charged particle, $-\frac{{dE}}{{dt}}$, where $E$ is the energy of the charged particle. We further recall that the $w$-integration runs over the frequency range for which $n(w) > {c \mathord{\left/
 {\vphantom {c v}} \right.\kern-\nulldelimiterspace} v}$. It is to be specially noted that the refractive index of the electrodynamics studied in this work has a rather geometrical origin than in a polarizable medium such as the QED vacuum, made up of virtual electrons and positrons. Incidentally, it will also be observed that we are in presence of a medium whose dispersive character has a geometrical origin. As already mentioned, the vacuum behaves as a new medium with an underlying quantum geometry with fundamental blocks of gravity that arise from the granularity of space-time. Having in mind that the vacuum is made up of virtual electrons and positrons,  we may set an upper bound for the integration over $w$ in (\ref{II-85}). Our analysis follows closely that of \cite{Cherenkov23}.                 
 
With this then, we pass now to consider the new geometrical vacuum analogous to the general case of nonlinear electrodynamics, where the polarizable medium is the QED vacuum, made up of virtual electrons and positrons. In such a case, the produced Cherenkov radiation corresponds to the energy re-emitted by the excited virtual particles. In fact, we have argued that the frequency, $\frac{2mc^2}{\hbar}$, acts as a cutoff frequency for the re-emitted photons, which then corresponds to the energy for a pair creation \cite{Cherenkov23}. This can be seen as follows, it is plausible to think that the charged particle is not able to excite virtual pairs in the QED vacuum with the energy necessary for the pair creation. Hence, the re-emitted energy must be upper-bounded by $2mc^{2}$ which corresponds to the cutoff  $\Omega  = \frac{2mc^2}{\hbar}$ on the frequencies to be integrated over. Evidently, this frequency corresponds to the energy for a pair creation.  

From this last remark it follows that equation (\ref{II-85}) then becomes 
\begin{eqnarray}
{\mathcal{E}} &=& \frac{{2}\mathit{\pi}{Q}^{2}}{{c}^{2}}\mathop{\int}\nolimits_{0}\nolimits^{\Omega}{dw}\,{w}\left({{1}{-}\frac{{v}^{2}}{{c}^{2}}}\right){-}\frac{{4}\mathit{\pi}{Q}^{2}\mathit{\eta}}{{c}^{2}}{l}_{P}\mathop{\int}\nolimits_{0}\nolimits^{\Omega}{dw}\,{w}^{2} \nonumber\\
&=&
\frac{{4}\mathit{\pi}\,{Q}^{2}\,{m}^{2}\,{c}^{2}}{{\hbar}^{2}}\left({{1}{-}\frac{{v}^{2}}{{c}^{2}}}\right){-}\frac{{32}\mathit{\pi}\,{Q}^{2}\mathit{\eta}\,{m}^{3}\,{c}^{4}}{3{\hbar}^{3}}{l}_{P},\label{II-90}
\end{eqnarray}
which explicitly illustrates the dependency on $l_P$.

\section{Final remarks}

In this work, based on the standard approach of calculating the Poynting vector, we have provided another observational signature of Gambini-Pullin electrodynamics.

In fact, we would like to justify the emergence of Cherenkov
radiation in the system we are considering. The extended
Maxwell-like system we are studying exhibits loop quantum gravity (LQG) effects
in the Faraday-Lenz and Ampère-Maxwell equations. They
are both linear and the medium is the quantum electrodynamics (QED) vacuum, so it
would appear that there should be no room for Cherenkov
radiation. However, the picture we should have in mind is
that the lighter-order space derivatives present in the Faraday-
Lenz and Amp\`ere-Maxwell equations stem from the geometry
that describes the granularity of space-time at the Planck scale.
So, the virtual electron-positron sector of the QED vacuum
probed by the electromagnetic waves supports Cherenkov 
radiation as a result of the underlying geometry dictated by
the LQG scenario at the Planck scale. Cherenkov radiation
is naturally produced in QED vacuum in the framework of
non-linear extensions of electrodynamics. Here, the Cherenkov
effect in QED vacuum happens by virtue of the granularity of
space-time.

As a follow-up of the investigation pursued in this contribution, we envisage to study possible consequences of non-linear terms – actually, cubic terms – in the magnetic field induced by the granular geometry of LQG in the Amp\`ere-Maxwell equation, as reported in the paper \cite {Li_Ma}. The refractive index, n, of eq. (\ref{II-85}) will no longer depend only on the frequency, but there will come out a dependence on a background magnetic field that might eventually be present. Since we are taking into account Planck scale effects, it is sensible to apply our results to an early Universe scenario. With this picture in mind, we shall analyze how the dependence of the refractive index on a strong external magnetic field of the early Universe interferes on the production of Cherenkov waves.

\section*{Acknowledgments}

One of us (P. G.) was partially supported by ANID PIA / APOYO AFB220004 (Chile).

\end{document}